\documentclass[12pt]{article}
\catcode`\@=11
\@addtoreset{equation}{section}

\global\arraycolsep=2pt 
\newcommand{\ga}{\gamma}

\newcommand{\la}{\lambda}

\newcommand{\Tr}{{\rm Tr}}

\newcommand{\alg}[1]{\mathfrak{#1}}

\newcommand{\mL}{\mathcal L}

\usepackage{mathrsfs,amsbsy,amssymb,latexsym,amsfonts,amsmath,cite}
\usepackage{graphicx,color}
\usepackage{mathtools}

\allowdisplaybreaks

\begin{document}

\begin{flushright}
\parbox{4cm}
{KUNS-2599 \\ 
\today }
\end{flushright}

\vspace*{1.5cm}

\begin{center}
{\Large \bf 
Lax pairs for deformed  Minkowski spacetimes 
}
\vspace*{1.5cm}\\
{\large Hideki Kyono\footnote{E-mail:~h\_kyono@gauge.scphys.kyoto-u.ac.jp}, 
Jun-ichi Sakamoto\footnote{E-mail:~sakajun@gauge.scphys.kyoto-u.ac.jp}
and Kentaroh Yoshida\footnote{E-mail:~kyoshida@gauge.scphys.kyoto-u.ac.jp}} 
\end{center}

\vspace*{0.5cm}

\begin{center}
{\it Department of Physics, Kyoto University, \\ 
Kitashirakawa Oiwake-cho, Kyoto 606-8502, Japan} 
\end{center}

\vspace{1cm}

\begin{abstract}
We proceed to study Yang-Baxter deformations of 4D Minkowski spacetime 
based on a conformal embedding. We first revisit a Melvin background and 
argue a Lax pair by adopting a simple replacement law invented in 1509.00173. 
This argument enables us to deduce a general expression of Lax pair. 
Then the anticipated Lax pair is shown to work for arbitrary classical $r$-matrices 
with Poincar\'e generators. As other examples, we present Lax pairs 
for pp-wave backgrounds, the Hashimoto-Sethi background, 
the Spradlin-Takayanagi-Volovich background.

\end{abstract}

\setcounter{footnote}{0}
\setcounter{page}{0}
\thispagestyle{empty}

\newpage

\tableofcontents

\section{Introduction}

The Yang-Baxter sigma-model description is known as a systematic way 
to study integrable deformations of 2D integrable non-linear sigma models \cite{Klimcik}. 
It was originally proposed for principal chiral models on the basis of the modified classical 
Yang-Baxter equation (mCYBE)\,. It was generalized to symmetric cosets 
by Delduc, Magro and Vicedo \cite{DMV} and then to the homogeneous classical 
Yang-Baxter equation (CYBE) \cite{MY-YBE}. 

\medskip 

Yang-Baxter deformations can be applied to integrable deformations of 
the AdS$_5 \times$S$^5$ superstring. The first example is a $q$-deformation 
of the AdS$_5\times$S$^5$ superstring \cite{DMV2} based on the mCYBE. 
This deformations is often called the $\eta$-deformation. 
The metric and NS-NS two-form are obtained in \cite{ABF}. 
The remaining fields are determined recently by performing a super coset construction 
at the level of quadratic fermions \cite{ABF2}. As a result, 
the obtained fields do not satisfy the on-shell condition of type IIB supergravity. 
(For another approach to an exact solution including T-duals of the metric, see \cite{HT-sol}). 
However, in the very recent paper \cite{scale}, it is shown that the scale invariance of 
the world-sheet theory is still preserved and it has been conjectured 
the background satisfies rather the modified type IIB supergravity 
equations of motion in which the RR-field strengths are of the second order in derivatives. 

\medskip

Another class of Yang-Baxter deformations of the AdS$_5\times$S$^5$ superstring 
is based on the homogeneous CYBE \cite{KMY-Jordanian-typeIIB}. 
The deformations include TsT transformations of AdS$_5\times$S$^5$ 
\cite{LM-MY,MR-MY,Sch-MY} such as $\gamma$-deformations of S$^5$ \cite{LM,Frolov}, 
gravity duals of noncommutative gauge theories \cite{HI,MR}, 
and Schr\"odinger geometries \cite{MMT}.
These backgrounds can also be realized as the undeformed AdS$_5\times$S$^5$ 
with twisted boundary conditions \cite{Frolov,AAF,KY-Sch,Stijn,Benoit,KKSY}. 
Other deformations are associated with classical $r$-matrices composed of 
non-commuting generators and lead to deformed backgrounds obtained through 
a chain of dualities including S-dualities \cite{MY-duality,KMY-SUGRA}. 
Further remarkably, these deformations may work for non-integrable backgrounds, 
such as a Sasaki-Einstein manifold $T^{1,1}$ \cite{BZ}. 
TsT transformations of $T^{1,1}$ \cite{LM,CO} are reproduced as Yang-Baxter deformations \cite{CMY}. 
Thus the connection between gravity solutions and classical $r$-matrices 
may deserve to be called the gravity/CYBE correspondence 
(For a short summary see \cite{MY-summary}).

\medskip

Furthermore, we have considered Yang-Baxter deformations of 4D Minkowski spacetime 
as a generalization, in order to examine applicability of the correspondence 
\cite{YB-Min, kappa-Min}\footnote{For an approach based on a scaling limit of 
$q$-deformed AdS$_5\times$S$^5$\,, see \cite{Pachol}.}. 
Firstly, the work \cite{YB-Min} showed that the Yang-Baxter deformations can reproduce 
twisted backgrounds such as Melvin backgrounds \cite{Melvin,GM,Tseytlin,HT}, 
pp-wave backgrounds \cite{NW} and time-dependent exactly-solvable backgrounds including 
the Hashimoto-Sethi background \cite{HS} and the Spradlin-Takayanagi-Volocich background \cite{STV},  
like in the case of AdS$_5\times$S$^5$\,. Then the analysis has been generalized 
by adopting classical $r$-matrices which describe $\kappa$-deformations of Poincar\'e algebra \cite{kappa}.
The associated deformed backgrounds include T-duals of (A)dS$_4$ 
and a time dependent pp-wave background. 
Interestingly, these backgrounds can also be reproduced 
from different classical $r$-matrices including the dilatation \cite{YB-Min}.
In addition, the Lax pair has been constructed for the general $\kappa$-deformations \cite{kappa-Min}.

\medskip

The construction of the Lax pair in \cite{kappa-Min} utilized the expression of 
Lax pair for arbitrary classical $r$-matrices 
which are composed of Poincar\'e generators and satisfy the CYBE, 
while the derivation of this part has not been explained. 
Our main purpose here is to provide the derivation 
as a follow-up of the previous work \cite{kappa-Min}. 
We first revisit a Melvin background and argue the associated Lax pair 
by following a simple replacement law invented in \cite{KKSY}. 
This argument enables us to deduce a general expression of Lax pair. 
Then we show that the anticipated Lax pair works for arbitrary classical $r$-matrices 
with Poincar\'e generators. As other examples, we present Lax pairs for pp-wave backgrounds, 
the Hashimoto-Sethi background, the Spradlin-Takayanagi-Volovich background. 

\medskip

This parer is organized as follows. In section 2, we give a brief review of Yang-Baxter deformations 
of 4D Minkowski spacetime. In section 3, a Melvin background is revisited  
and the associated Lax pair is constructed by following a simple replacement rule invented in \cite{KKSY}. 
A twisted boundary condition is also discussed. 
In section 4, we show that the anticipated Lax pair works for arbitrary classical $r$-matrices which are
composed of Poincar\'e generators. 
In section 5, we present Lax pairs and argue twisted boundary conditions 
for the following examples: 1) pp-wave backgrounds, 
2) the Hashimoto-Sethi background, 3) the Spradlin-Takayanagi-Volovich background. 
Section 6 is devoted to conclusion and discussion. 
In Appendix A, our notation and convention are summarized. 
Appendix B gives a short explanation of a conformal embedding of 4D Minkowski spacetime.

\section{Yang-Baxter deformations of Minkowski spacetime}\label{sec:YB-4D-Min}

We shall present a short review of Yang-Baxter deformations of 
4D Minkowski spacetime by following the previous works \cite{YB-Min,kappa-Min}. 
For our notation and convention, see Appendix A. 

\medskip 

A significant point is that the problem of the degenerate Killing form 
can be avoided by adopting a conformal embedding as briefly described in Appendix B. 
In the following, the string tension $T =\frac{1}{2\pi\alpha'}$ is set to 1, 
and the conformal gauge is taken so as to drop the dilaton coupling 
to the world-sheet scalar curvature. 

\medskip 

The Yang-Baxter deformed action is given by \cite{YB-Min}
\begin{equation}
\label{action}
S=-\frac{1}{2}\int_{-\infty}^{\infty}\!\!d\tau\int_{0}^{2\pi}\!\!d\sigma\,
(\gamma^{\alpha\beta}-\epsilon^{\alpha\beta})\, 
\Tr\Biggl[A_{\alpha}P\circ\frac{1}{1-2\eta R_{g}\circ P}\,A_{\beta}\Biggr]\,. 
\end{equation}
The left-invariant one-form $A_{\alpha} \equiv g^{-1}\partial_{\alpha}g$ is defined with a group element  
$g$ given in (\ref{para})\,. The projection $P$ is defined as 
\begin{align}
P(x) \equiv \frac{1}{4} \Bigl[-\ga_{0}\,\Tr(\ga_{0}\,x) + \sum_{i=1}^3\ga_{i}\, \Tr(\ga_{i}\,x) \Bigr] 
\label{P}
\end{align}
and is closely related to the coset structure of AdS$_5$\,. 
The deformation is measured by $\eta$\,. The undeformed action is reproduced when $\eta=0$\,. 
The world-sheet metric is given by $\gamma_{\alpha\beta}={\rm diag}(-1,1)$\,. 
The anti-symmetric tensor $\epsilon^{\alpha\beta}$ is normalized as $\epsilon^{\tau\sigma}=1$\,. 

\medskip

A key ingredient is a linear $R$-operator contained  in $R_g$ defined as 
\begin{equation}
R_g \equiv g^{-1}R(g Xg^{-1})g\,.
\label{min-R}
\end{equation}
The linear operator $R : \alg{so}(2,4)\to \alg{so}(2,4)$ is a solution of the mCYBE \cite{mCYBE}\,,
\begin{equation}
\bigl[R(x),R(y)\bigr]-R\left([R(x),y]+[x,R(y)]\right)=\omega\, [x, y]\,,\qquad x, y\in \alg{so}(2,4)\,,
\label{CYBE-2}
\end{equation}
where $\omega$ is a constant parameter. 
The $R$-operator corresponds to a {\it skew-symmetric} classical $r$-matrix 
in the tensorial notation through the following formula:
\begin{equation}
R(X)=\Tr_{2}[r(1\otimes X)]=\sum_{i}(a_{i}\Tr(b_{i}X)-b_{i}\Tr(a_{i}X))\,.  \label{rel}
\end{equation}
Here the classical $r$-matrix is represented by 
\begin{equation}
r=\sum_{i}a_{i}\wedge b_{i}\equiv \sum_{i}(a_{i}\otimes b_{i}-b_{i}\otimes a_{i})\,.
\end{equation}
The generators $a_{i}, b_{i}$ are elements of $\alg{so}(2,4)$\,. 
This means that Yang-Baxter deformations are investigated based 
on the extended algebra $\alg{so}(2,4)$\,, rather than the Poincar\'e algebra. 

\medskip 

In particular, when $\omega=0$\,, the equation (\ref{CYBE-2}) is called the homogenous CYBE.
When (skew-symmetric) classical $r$-matrices satisfy the standard CYBE, 
\begin{equation}
\left[r_{12},r_{13}\right] +\left[r_{12},r_{23}\right]+\left[r_{13},r_{23}\right]=0\,, \label{CYBE-1}
\end{equation}
they lead to $R$-operators which satisfy the homogenous CYBE. In the following, we are mainly 
concerned with the homogeneous CYBE case.

\medskip

To construct Lax pairs, it is useful to work with the following deformed current, 
\begin{eqnarray}
J_\pm=\frac{1}{1\mp2\eta R_g\circ P}(A_\pm)\,.
\end{eqnarray}
It is also helpful to introduce the world-sheet light-cone components as  
\[
A_\pm\equiv A_\tau\pm A_\sigma\,.
\] 
Then the Lagrangian in (\ref{action}) can be rewritten into a simpler form: 
\begin{eqnarray}
L=\frac{1}{2}\Tr[A_-P(J_+)]\,.
\end{eqnarray}
One can read off the deformed metric and NS-NS two-form 
from the symmetric and skew symmetric parts 
with respect to the world-sheet coordinates 
in the deformed Lagrangian, respectively.

\medskip 

It should be remarked that, although it is sufficient to compute an explicit form 
of the projected deformed current $P(J)$
to study the deformed metric, the explicit form of the deformed current $J$ itself  
is necessary to construct Lax pairs.

\subsection*{A schematic classification of classical $r$-matrices}

It would be helpful to list up possible classical $r$-matrices concerned with our analysis. 
Schematically, those are classified into the following three classes: 
\begin{enumerate}
\item[(a)] \quad $r=$ Poincar\'e $\otimes$ Poincar\'e \\ 
\quad 1. ~~ abelian ~~e.g., \quad $r \sim p_{1} \wedge p_{2}$\,, \qquad 
2. ~~ non-abelian ~~ e.g., \quad $r \sim p_{1} \wedge n_{12}$\,, 
\item[(b)] \quad $r=$ Poincar\'e $\otimes$ non-Poincar\'e \\ 
\quad 1. ~~ abelian ~~e.g., \quad $r \sim n_{12} \wedge \hat{d}$\,, \qquad 
2. ~~ non-abelian ~~e.g., \quad $r \sim p_0 \wedge \hat{d}$\,,   
\item[(c)] \quad $r=$ non-Poincar\'e $\otimes$ non-Poincar\'e \\ 
\quad 1. ~~ abelian ~~e.g., \quad $r \sim k_{1} \wedge k_{2}$\,, \qquad 
2. ~~ non-abelian ~~e.g., \quad $r \sim k_0 \wedge \hat{d}$\,.  
\end{enumerate}
Note that, given a classical $r$-matrix $r = a\otimes b$\,,  
the word (non-)abelian means that $a$ and $b$ (do not) commute each other. 
The generators of 4D conformal algebra $p_\mu\,,n_{\mu\nu}\,,k_\mu$ and $\hat{d}$ 
are associated with translations, Lorentz rotations, special conformal transformations and dilatation, respectively 
(For our notation and convention, see Appendix A).

\medskip 

The complete classification of classical $r$-matrices has not been done and 
the associated Yang-Baxter deformations have also not been investigated completely. 
It is nice to look for the relation between classical $r$-matrices and deformed geometries 
furthermore. For classical $r$-matrices in the class (a)\,, 
it would be helpful to consult an intensive classification in \cite{Tolstoy,Z}.  

\medskip 

So far, classical $r$-matrices in the classes (a)-1 and (b)-2 have been studied in \cite{YB-Min,kappa-Min}. 
The class (a)-1 is intimately related to (generalized) Melvin twists \cite{YB-Min}. 
Some examples of the class (b)-2 have been studied as well. 
In particular, T-duals of dS$_4$ and AdS$_4$ are realized with 
$r \sim p_0 \wedge \hat{d}$ and $r \sim p_1 \wedge \hat{d}$\,, respectively. 
Interestingly, special classical $r$-matrices in the class (a)-2, which are also used 
in the context of $\kappa$-deformations of Poincar\'e algebra, 
lead to the identical T-duals of dS$_4$ and AdS$_4$ \cite{kappa-Min}.  

\medskip 

Although we have succeeded in constructing a Lax pair for the general $\kappa$-deformations, 
the construction was based on the expression of Lax pair for the class (a) with the homogeneous CYBE 
and we have not derived it explicitly. 

\medskip 

In the following, we will present a method to construct Lax pairs 
for the classical $r$-matrices in the class (a) that satisfy the homogeneous CYBE.   
A simple replacement law based on TsT transformations is utilized to deduce a general form of Lax pair 
for arbitrary classical $r$-matrices in the class (a)-1. 
Finally, the anticipated Lax pair is shown to work for all classical $r$-matrices in the class (a) 
with the homogeneous CYBE. It is straightforward to generalize the general form to the mCYBE case, 
by following the previous work \cite{kappa-Min}.

\section{A Melvin background revisited}

Let us revisit a Melvin background here. We derive a Lax pair by applying a simple replacement 
rule to the one for 4D Minkowski spacetime.  
Then the result enables us to deduce a general form of Lax pair for an arbitrary classical $r$-matrix 
of the class (a)-1. We also argue a general form of twisted boundary condition with which the undeformed 
background is equivalent to the deformed background.

\subsection{Lax pair for 4D Minkowski spacetime}

We first construct the Lax pairs for sigma models on 4D Minkowski spacetime. 
The starting classical action can be expressed as 
\begin{equation}
S=\frac{1}{2}\int_{-\infty}^{\infty}\!\!d\tau\int_{0}^{2\pi}\!\!d\sigma\,
\Tr\bigl[A_-P(A_+)\bigr]\,.
\end{equation}
Then the associated Lax pair is given by
\begin{eqnarray}
\mL_\pm^{\text{Min}}=P_0(A_\pm)+\la^{\pm1}P_2(A_\pm)
\label{Min-Lax}
\end{eqnarray}
with a spectral parameter $\lambda\in \mathbb{C}$\,. 
Here $P_0$ and $P_2 \equiv P+P'$ 
are the projection operators, where $P$ is already introduced in (\ref{P}), 
and $P_0$ and $P'$ are defined as, respectively, 
\begin{eqnarray}
P_0(x) \equiv \frac{1}{2}\sum_{\mu,\nu=0}^3n_{\mu\nu}\, 
\frac{\Tr(n_{\mu\nu}x)}{\Tr(n_{\mu\nu}n_{\mu\nu})}\,,\qquad
P'(x) \equiv \sum_{\mu=0}^3n_{\mu5}\, \frac{\Tr(n_{\mu5}x)}{\Tr(n_{\mu5}n_{\mu5})}\,.
\end{eqnarray}
For the standard coset representative (\ref{para}), it is easy to check that 
the equations of motion are equivalent to the flatness condition of the Lax pair (\ref{Min-Lax}).
Note that $P_0$ and $P_2$ are regarded as the projections to the $\mathbb{Z}_2$-graded components 
of 4D Poincar\'e algebra $\mathfrak{iso}(1,3)$\,, 
\begin{eqnarray}
P_0&:& \quad \mathfrak{iso}(1,3) ~~\longrightarrow~~ 
\mathfrak{iso}^{(0)}(1,3) = \mathfrak{so}(1,3)\,, \nonumber \\
P_2&:& \quad \mathfrak{iso}(1,3) ~~\longrightarrow~~ 
\mathfrak{iso}^{(2)} (1,3) = \mathfrak{iso}(1,3)/\mathfrak{so}(1,3)\,. 
\end{eqnarray}  

\subsection{Lax pair for a Melvin background}

Next, we will derive a Lax pair for a Melvin background \cite{Melvin, GM, Tseytlin, HT} 
associated with an abelian classical $r$-matrix 
\begin{eqnarray}
r=\frac{1}{2}\,p_3\wedge n_{12}\,.
\end{eqnarray}
This $r$-matrix leads to the metric, 
\begin{eqnarray}
ds^2 &=& -(dx^0)^2+dr^2+G_{\rm M}\left[\,r^2d\theta^2+ (dx^3)^2\,\right]\,,\nonumber \\
B &=& \eta\, r^2\, G_{{\rm M}}\,d\theta\wedge dx^3\,,\qquad
G^{-1}_{{\rm M}} \equiv 1+\eta^2 r^2\,, \label{Melvin}
\end{eqnarray}
with the standard coset representative (\ref{para}) and the polar coordinates
\begin{eqnarray}
x^1=r\cos\theta\,,\qquad x^2=r\sin\theta\,.
\end{eqnarray}
This background describes an example of Melvin backgrounds. 

\medskip 

Now that the left-invariant one-form $A$ can be expanded as 
\begin{eqnarray}
A_\pm &=& \partial_\pm x^0\,p_0 + \bigl[\cos\theta\,\partial_\pm r 
-r\sin\theta\,\partial_\pm\theta \bigr]\,p_1 \nonumber \\ 
&&  +\bigl[\sin\theta\,\partial_\pm r + r \cos\theta\,\partial_\pm\theta\bigr]\,p_2 
+\partial_\pm x^3\,p_3 \,, 
\end{eqnarray}
the deformed current $J$ is evaluated as 
\begin{eqnarray}
J_\pm&=&\partial_\pm x^0\,p_0+G_{{\rm M}}\,(\partial_\pm x^3 \mp \eta\, r^2\,\partial_\pm \theta)
\,(p_3-\eta\,n_{12})\nonumber \\
&&+\left[\cos\theta\,\partial_\pm r 
- G_{\text{M}}\,r\sin\theta\,(\partial_\pm\theta\pm\eta\,\partial_\pm x^3)\right]p_1\nonumber \\
&&+\left[\sin\theta\,\partial_\pm r
+ G_{\text{M}}\,r\cos\theta\,(\partial_\pm\theta\,\pm\eta\,\partial_\pm x^3)\right]p_2\,.
\label{dcurrent}
\end{eqnarray}
Comparing $J$ with $A$\,, one can find the following replacement rule:
\begin{eqnarray}
\partial_\pm\theta \quad &\longrightarrow& \quad 
G_{{\rm M}}\,\bigl(\partial_\pm\theta\pm\eta\, \partial_\pm x^3 \bigr)\,,\nonumber \\
\partial_\pm x^3 \quad &\longrightarrow& \quad 
G_{{\rm M}}\, \bigl(\partial_\pm x^3\mp \eta\, r^2\,\partial_\pm \theta \bigr)\,.
\label{Melvin-TsT}
\end{eqnarray}
Note that this rule makes sense at the off-shell level. 

\medskip 

One can derive a Lax pair for the Melvin background (\ref{Melvin}) 
by applying the replacement rule (\ref{Melvin-TsT}) to the undeformed one. 
First of all, it is necessary to prepare the undeformed Lax pair that has a manifest $U(1)$ isometry 
along the $\theta$-direction.  
It is easy to remove the $\theta$-dependence of the undeformed Lax 
by performing a gauge transformation,  
\begin{eqnarray}
\mL_\pm^{\text{Min}} ~~\to~~ \mL_\pm^{\text{Min},h}
&=&h^{-1}\mL_\pm^{\text{Min}} h+h^{-1}\partial_\pm h\nonumber \\
&=&-\partial_\pm \theta\,n_{12}+\la^{\pm1}(\partial_\pm r\,p_1
+r\partial_\pm \theta\,p_2+\partial_\pm x^3\,p_3+\partial_\pm x^0\,p_0) 
\label{gaugetr}
\end{eqnarray}
with $h=\exp(-\theta\,n_{12})$\,. 
The resulting Lax pair depends only on the derivative of $\theta$ 
and hence has a $U(1)$ symmetry along the $\theta$-direction. 
Thus, by applying the replacement rule (\ref{Melvin-TsT}) to (\ref{gaugetr})\,, 
the Lax pair for the Melvin background is given by 
\begin{eqnarray}
\mL^{\text{Melvin}}_\pm&=&-G_{{\rm M}}(\partial_\pm\theta\pm\eta \partial_\pm x^3)\,n_{12}
+\la^{\pm1}\bigl[\partial_\pm x^0\,p_0+\partial_\pm r\,p_1\nonumber \\
&&\qquad\qquad+G_{{\rm M}}\,r(\partial_\pm\theta\pm\eta \partial_\pm x^3)\,p_2
+G_{{\rm M}}(\partial_\pm x^3\mp\eta r^2\partial_\pm \theta)\,p_3\bigr]\,. 
\label{Lax-Melvin}
\end{eqnarray}
One can readily check that the zero-curvature condition of this Lax pair 
is equivalent to the equations of motion for the Melvin background.

\medskip

Now we can deduce a general form of Lax pair for arbitrary classical $r$-matrices in the class (a)-1\,. 
First, with the deformed current $J$ in (\ref{dcurrent})\,, 
the Lax pair (\ref{Lax-Melvin}) can be rewritten into the following form: 
\begin{eqnarray}
\mL^{\text{Melvin}}_\pm = h^{-1}\partial_\pm h+h^{-1}P_0(J_\pm)h+\la^{\pm1}h^{-1}P_2(J_\pm)h\,.
\end{eqnarray}
The flatness condition of the Lax pair should be preserved under arbitrary gauge transformations. 
Thus the resulting Lax pair is given by 
\begin{eqnarray}
\mL^{\rm Melvin}_\pm=P_0(J_\pm)+\la^{\pm1}P_2(J_\pm)\,.
\label{Def-Lax}
\end{eqnarray}

\medskip 

This abstract expression is really significant because one can anticipate that 
the form (\ref{Def-Lax}) would hold for arbitrary classical $r$-matrices in the class (a)-1  
satisfying the homogeneous CYBE. In fact, this anticipation is true as proven in the next section. 
More interestingly, the Lax pair (\ref{Def-Lax}) works for the class (a)-2 as well. 
Thus, in total, the Lax pair (\ref{Def-Lax}) holds for arbitrary classical $r$-matrices in the class (a) 
satisfying the homogeneous CYBE.

\subsection{Twisted boundary condition}

The Melvin background (\ref{Melvin}) can be realized 
as a Melvin twist of 4D Minkowski spacetime \cite{Melvin}. 
We will reconsider here the Melvin background (\ref{Melvin}) 
from the point of view of a twisted boundary condition for 4D Minkowski spacetime. 

\medskip 

For this purpose, it is useful to take the following coset representative: 
\begin{eqnarray}
g(\tau,\sigma)&=&\exp\left[x^3\,p_3-\theta\,n_{12}\right]\exp\left[x^0\,p_0+r\,p_1\right]\nonumber \\
&=&\exp[x^0\,p_0+r\cos\theta\,p_1+r\sin\theta\,p_2+ x^3\,p_3]\exp[-\theta\,n_{12}]
\label{M-para}\,.
\end{eqnarray}
This representative (\ref{M-para}) is equivalent to the standard one (\ref{para}) 
under the gauge transformation (\ref{gaugetr})\,. 

\medskip

Then, with the replacement rule (\ref{Melvin-TsT})\,, we will show that 
the deformed sigma model can be mapped to the undeformed one with a twisted boundary condition. 
In the first place, one needs to obtain $U(1)$ currents in the undeformed 
and deformed backgrounds, respectively,
\begin{eqnarray}
\tilde{\mathcal{J}_\theta^\alpha} &=& -\ga^{\alpha\beta}\, r^2\,\partial_\beta\tilde{\theta}\,,\nonumber \\
\tilde{P}_3^{\alpha} &=& -\ga^{\alpha\beta}\, \partial_\beta \tilde{x}^3\,,\nonumber \\
\mathcal{J}_\theta^\alpha &=& -\ga^{\alpha\beta} \,G_M\, 
r^2\, \bigl[\partial_\beta\theta-\eta\,\ga_{\beta\ga}\,\epsilon^{\ga\rho}\partial_\rho x^3\bigl]\,, \nonumber \\
P_3^\alpha&=&-\ga^{\alpha\beta} \,G_M 
\bigl[\partial_\beta x^3+\eta\,\ga_{\beta\ga}\,\epsilon^{\ga\rho}r^2\partial_\rho\theta \bigr]\,.
\end{eqnarray}
Here $\tilde{x}^3$ and $\tilde{\theta}$ represent coordinates on the undeformed background.
With the rule (\ref{Melvin-TsT}), one can show that the $U(1)$ currents 
in 4D Minkowski spacetime are equal to those in the Melvin background, i.e.,
\begin{eqnarray}
\tilde{\mathcal{J}}_{\theta}^{\alpha} = \mathcal{J}_{\theta}^{\alpha}\,,\qquad 
\tilde{P}_3^{\alpha} = P_3^{\alpha}\,.
\label{current}
\end{eqnarray}
The temporal components of the relations in (\ref{current}) 
imply that $P^\tau_3$ and $\mathcal{J}_\theta^\tau$ 
are equivalent to $\tilde{\mathcal{J}}_\theta^\tau$ and $\tilde{P}_3^\tau$\,. 
Then, by integrating the spacial components of the relations in (\ref{current}), 
twisted boundary conditions for the undeformed coordinates 
$\tilde{\theta}$ and $\tilde{x}^3$ are obtained as 
\begin{eqnarray}
\tilde{\theta}(\tau,2\pi)&=&\tilde{\theta}(\tau,0)+\eta\,P_3+2\pi w\,,\nonumber \\
\tilde{x}^3(\tau,2\pi)&=&\tilde{x}^3(\tau,0)-\eta\,\mathcal{J}_\theta\,. \label{tbc}
\end{eqnarray}
Here $w\in \mathbb{Z}$ is a winding number for the angular coordinate $\theta$\,.  
The conserved charges $P_3$ and $\mathcal{J}_\theta$ are defined as 
\begin{eqnarray}
P_3\equiv \int_0^{2\pi}\!\!d\sigma\,P_3^\tau\,,\qquad 
\mathcal{J}_\theta\equiv \int_0^{2\pi}\!\!d\sigma\,\mathcal{J}_\theta^\tau\,.
\end{eqnarray}

\medskip 

Note here that there is a relation between a gauge transformed current $J$ 
\[
J^g_\alpha=gJ_\alpha g^{-1}+g\partial_\alpha g^{-1} 
\]
and the $U(1)$ currents
\begin{eqnarray}
J_\sigma^g&=&-\eta r^2G_{{\rm M}}(\dot{\theta}+\eta x'^3)p_3
-\eta G_{{\rm M}}(\dot{x}^3-\eta r^2\theta')n_{12}\nonumber \\
&=&-\eta\, \mathcal{J}_\theta^\tau\, p_3 - \eta\, P_3^\tau\, n_{12}\,.
\end{eqnarray}
Thus the twisted boundary conditions (\ref{tbc}) can be recast into the following form:
\begin{eqnarray}
\tilde{g}(\tau,2\pi)=\text{P}\exp\left[\int_0^{2\pi}J^g_\sigma\,d\sigma\right]
\exp\Bigl( -2\pi\, w\,n_{12} \Bigr)\,\tilde{g}(\tau,0)\,.
\label{M-twist}
\end{eqnarray}
Here $\tilde{g}(\tau,\sigma)$ is parametrized as in (\ref{M-para}) 
with the undeformed coordinates ($\tilde{x}^0, \tilde{r}, \tilde{\theta}, \tilde{x}^3$)\,.

\section{General arguments}

In this section, let us show that the anticipated form of Lax pair
in Sec.\,3.2 works for arbitrary classical $r$-matrices of the class (a) satisfying the CYBE. 
Then we consider a general twisted boundary condition for the class (a)-1. 
Finally, we argue a general expression of Lax pair for the mCYBE case. 

\subsection{Proof of the anticipated Lax pair}

We show that the flatness condition of the anticipated Lax pair (\ref{Min-Lax}) 
is equivalent to the equations of motion and 
confirm that the Lax pair (\ref{Min-Lax}) works well 
for all classical $r$-matrices that are solutions of the homogeneous CYBE. 

\medskip 

It is convenient to rewrite the Lax pair (\ref{Min-Lax}) as
\begin{eqnarray}
\label{hLax}
\mL_\pm&=&J^n_\pm+\lambda^{\pm1}J^p_\pm\,, \qquad 
J^p_\pm \equiv J^{\mu}_\pm p_\mu\,,\quad J^n_\pm 
\equiv J^{\mu\nu}_\pm n_{\mu\nu}\,.
\end{eqnarray}
The zero-curvature condition of the Lax pair (\ref{Min-Lax}) can also be rewritten as
\begin{eqnarray}
\label{lax zcc}
0&=&\partial_+\mL_--\partial_-\mL_++\left[\mL_+,\mL_-\right] \nonumber\\
&=&\lambda\left(-\partial_-J^p_++[J^p_+,J^n_-]\right)
+\frac{1}{\lambda}\left(\partial_+J^p_-+[J^n_+,J^p_-]\right)  \nonumber\\
&&+\partial_+J^n_--\partial_-J^n_++[J^n_+,J^n_-]\,.
\end{eqnarray}
The coefficients of $\lambda^{\pm}$ and $\lambda^0$ should vanish respectively. 
Hence the zero-curvature condition is equivalent to the following set of three equations: 
\begin{eqnarray}
&& \partial_-J^p_+-[J^p_+,J^n_-]=0\,,\qquad \partial_+J^p_-+[J^n_+,J^p_-]=0\,,\nonumber\\
&& \partial_+J^n_--\partial_-J^n_+ + [J^n_+,J^n_-]=0\,.
\label{eom zcc}
\end{eqnarray}
Then the remaining task is to confirm that the three equations in (\ref{eom zcc}) 
are equivalent to the equations of motion of the deformed system with classical $r$-matrices 
of the class (a) 
and the zero-curvature condition for $A_\mu$\,.

\subsubsection*{Equations of motion}

Taking a variation of the classical action (\ref{action})\,, one can derive 
the equations of motion\footnote{The derivation of (\ref{eom}) is given in Appendix A of \cite{MY-YBE}. }
\begin{eqnarray}
\label{eom}
\Tr \bigl[\mathcal{E}\,p_\mu \bigr]=0\,,\quad\mathcal{E} 
\equiv \partial_+P(J_-)+\partial_-P(J_+)+[J_+,P(J_-)]+[J_-,P(J_+)]\,.
\end{eqnarray}
Then the equations in (\ref{eom}) can be rewritten as\footnote{Note that 
$J$ can be expanded with $p_{\mu}$ and $n_{\mu\nu}$ within the class (a)\,. } 
\begin{eqnarray}
\Tr \biggl[\bigl(\partial_+J^\rho_-\gamma_\rho+\partial_-J^\rho_+\gamma_\rho+
\left[J^\rho_+p_\rho+J^{\rho\sigma}_+n_{\rho\sigma},J^\lambda_-\gamma_\lambda\right]
+[J^\rho_-p_\rho+J^{\rho\sigma}_-n_{\rho\sigma}, 
J^\rho_+\gamma_\rho]\bigr)p_\mu \biggr]=0\,. 
\nonumber
\end{eqnarray}
In the third and fourth terms, the commutator 
$[\gamma_\mu,p_\nu]~(=2n_{\mu\nu}+\eta_{\mu\nu}\gamma_5)$ 
can be dropped off because the generators given 
by this commutator ($n_{\mu\nu}$ and $\gamma_5$) 
vanish after taking trace with $p_\mu$\,. 
For the remaining terms, $\gamma_\mu$ can be replaced 
by $p_\mu$ because the commutation relations $[\gamma_\mu,n_{\nu\rho}]$ 
and $[p_\mu,n_{\nu\rho}]$ take the same form. 
As a result, the equations of motion in (\ref{eom}) are equivalent to the following equations: 
\begin{eqnarray}
\label{eom2}
\tilde{\mathcal{E}}\equiv\partial_+J^p_-+\partial_-J^p_++[J^n_+,J^p_-]+[J^n_-,J^p_+]=0\,.
\end{eqnarray}

\subsubsection*{Zero-curvature condition}

One can rewrite the zero-curvature condition for $A_\pm$ in terms of $J_\pm$ like 
\begin{eqnarray}
\label{zcc}
0&=&\mathcal{Z}=\partial_+A_--\partial_-A_++\left[A_+,A_-\right] 
\nonumber\\
&=&\partial_+J_--\partial_-J_++\left[J_+,J_-\right]
+2\eta R_g(\mathcal{E})+4\eta^2{\rm{YBE}}_{Rg}(P(J_+),P(J_-))\,, \label{4.6}
\end{eqnarray}
through the relation $A_\pm=(1\mp 2\eta R_g\circ P)J_\pm$\,. 
Here the symbol ${\rm{YBE}}_{Rg}$ in the fifth term is defined as  
\begin{eqnarray}
{\rm{YBE}}_{Rg}(X,Y) \equiv [R_g(X),R_g(Y)]-R_g([R_g(X),Y]+[X,R_g(Y)])\,.
\end{eqnarray}
We are now considering the homogeneous CYBE case, hence the fifth term vanishes. 

\medskip 

Then, with the help of the relation 
\begin{equation}
R_g(\mathcal{E}) = R_g(\tilde{\mathcal{E}})\,, 
\end{equation}
the fourth term in the last line of (\ref{4.6}) can be rewritten as $2\eta R_g(\tilde{\mathcal{E}})$\,. 
Thus the last line of (\ref{zcc}) is recast into 
\begin{eqnarray}
\label{zcc2}
0&=&\mathcal{Z}=\partial_+J_--\partial_-J_++[J_+,J_-] 
+ 2\eta R_g (\tilde{\mathcal{E}})\,.
\end{eqnarray}
This equality can be decomposed into the $\mathfrak{iso}(1,3)/\mathfrak{so}(1,3)$ 
and $\mathfrak{so}(1,3)$ components like 
\begin{eqnarray}
\label{zcc3}
0&=&\mathcal{Z}^p\equiv\partial_+J^p_--\partial_-J^p_+ 
+ \left[J^p_+,J^n_-\right]+\left[J^n_+,J^p_-\right]+2\eta P(R_g(\tilde{\mathcal{E}}))\,,\nonumber\\
0&=&\mathcal{Z}^n\equiv\partial_+J^n_--\partial_-J^n_+ 
+ \left[J^n_+,J^n_-\right]+2\eta P_0(R_g(\tilde{\mathcal{E}}))\,.
\end{eqnarray}
Note that $P_0(R_g(\tilde{\mathcal{E}}))$ and $P(R_g(\tilde{\mathcal{E}}))$ 
vanish because $\tilde{\mathcal{E}}=0$\,, and this decomposition is based 
on the assumption that classical $r$-matrices are limited to the class (a). 

\subsubsection*{Equivalence}

Now we have prepared to check the equivalence between 
$\tilde{\mathcal{E}}=\mathcal{Z}^p=\mathcal{Z}^n=0$ 
and the equations in (\ref{eom zcc})\,. The manifest correspondence is summarized below: 
\begin{eqnarray}
\left\{
\begin{array}{c}
~~\tilde{\mathcal{E}}+\mathcal{Z}^p = 0 \\
~~\tilde{\mathcal{E}}-\mathcal{Z}^p = 0 \\
~~\mathcal{Z}^n = 0
\end{array}
\right.
\qquad \Longleftrightarrow \qquad \mbox{three equations in}~(\ref{eom zcc})\,.
\nonumber
\end{eqnarray}
Thus we have shown that the zero-curvature condition 
of the Lax pair (\ref{Min-Lax}) is equivalent to 
the equations of motion of the deformed system with classical $r$-matrices of the class (a) 
satisfying the homogeneous CYBE.

\subsection{Twisted boundary condition}

Let us next consider the general form of twisted boundary condition (\ref{M-twist})\,, 
which was anticipated in Sec.\,3.3, in the case of abelian classical $r$-matrices [class (a)-1]. 

\medskip 

Let us parametrize a group element $g$ as 
\begin{eqnarray}
g&=&g_A \, g_X\,,  \nonumber\\
g_A&=&\exp(x^a \,a+x^b\,b)\,,\quad g_X=\prod_i \exp(y^i\, X_i) 
\qquad \Bigl[ ~a\,,b\,,X_i\,\in\mathfrak{iso}(1,3)~\Bigr]\,, 
\label{Apara}
\end{eqnarray}
and suppose that a classical $r$-matrix composed of the generators 
$a$ and $b$ which appear in the definition of $g_A$
\begin{eqnarray}
r=a\wedge b\,,\qquad [a,\,b]=0\,.
\end{eqnarray}
Note here that this parameterization (\ref{Apara}) is assumed to be equivalent 
to the standard one (\ref{para}) via a gauge transformation like  
 \begin{eqnarray}
g({\rm \ref{para}})\rightarrow g({\rm \ref{para}})\,h\,,\quad h\in\mathfrak{so}(1,3)\,.
\end{eqnarray}
This parameterization\footnote{This parametrization has already been utilized in (\ref{M-para}) 
and will be used also in Sec.\ 5.} 
enables us to realize manifestly the translational invariance of the current $A_\alpha$ 
in the $x^a$ and $x^b$ directions. It is also easy to figure out   
a deformed part as a replacement law.
Therefore we will adopt the parameterization (\ref{Apara}) to argue a general form of 
twisted boundary condition. 

\subsubsection*{Replacement law}

Then the left-invariant one-form $A_\alpha =g^{-1}\partial_\alpha g$ 
is rewritten as 
\begin{eqnarray}
\label{UDC}
A_\alpha=\partial_\alpha x^a \,g^{-1}_X a\, g_X + \partial_\alpha x^b \,g^{-1}_X b\, 
g_X + g^{-1}_X \partial_\alpha g_X\,,
\end{eqnarray}
and the deformed current $J_\pm$ is evaluated as 
\begin{eqnarray}
\label{DC}
J_\pm&=&A_\pm\pm2\eta R_g\circ P(J_\pm)\nonumber\\
&=&(\partial_\pm x^a\pm2\eta\Tr[g_X^{-1}b\,g_X P(J_\pm)])\,g_X^{-1}a\,g_X\nonumber\\
&&+(\partial_\pm x^b\mp2\eta\Tr[g_X^{-1}a\,g_X P(J_\pm)])\,g_X^{-1}b\,g_X+g_X^{-1}\partial_\pm g_X\,.
\end{eqnarray}
Then, comparing (\ref{UDC}) with (\ref{DC})\,, one can read off the replacement law: 
\begin{eqnarray}
\label{RL}
\partial_\pm x^a  ~~&\rightarrow&~~ \partial_\pm x^a\pm2\eta\,\Tr[g_X^{-1}b\,g_X P(J_\pm)]\,,\nonumber\\
\partial_\pm x^b ~~&\rightarrow&~~ \partial_\pm x^b\mp2\eta\,\Tr[g_X^{-1}a\,g_X P(J_\pm)]\,.
\end{eqnarray}

\subsubsection*{Twisted boundary condition}

First of all, we show that the deformed action is invariant under the translations 
along the $x^a$ and $x^b$ directions. Then 
the second terms in the r.h.s.\ of (\ref{RL}) correspond to 
the Noether currents associated with these translations.

\medskip 

Let us consider a variation of the deformed action associated with 
an infinitesimal translation of $x^a$ and $x^b$
\begin{eqnarray}
x^a  ~~&\rightarrow&~~ x^a+\delta x^a\,,\qquad
x^b ~~\rightarrow~~ x^b+\delta x^b\,.
\end{eqnarray}
The associated variations of $g$ and $A_{\pm}$ are given by 
\begin{eqnarray}
\delta g=\epsilon\, g\,,\qquad\delta A_\pm=g^{-1}(\partial_\pm \epsilon )\, g\,.
\end{eqnarray}
Here an infinitesimal quantity $\epsilon$ is defined as  
\[
\epsilon ~\equiv~ \delta x^a\, a+\delta x^b\,b\,. 
\]
Then the variation of the deformed action is evaluated as 
\begin{eqnarray}
\delta S &=&-\int\! d^2\sigma\, \gamma^{\alpha\beta}\,{\rm Tr} 
\Bigl[\partial_\alpha\epsilon\bigl( gP(J_\beta)g^{-1})\Bigr]\nonumber\\
&=&-\int\! d^2\sigma\, \gamma^{\alpha\beta}\left(\partial_\alpha \delta x^a \,
\Tr \Bigl[g_X^{-1}a\,g_X\,P(J_\beta)\Bigr]
+\partial_\alpha \delta x^b\, \Tr \Bigl[g_X^{-1}b\,g_X\,P(J_\beta)\Bigr]\right)\,,
\label{VOA}
\end{eqnarray}
by using the following formula in taking a variation of $R_g$\,,  
\begin{eqnarray}
\delta R_g(X) &=& R_g (\delta X)+[R_g(X),g^{-1}\epsilon g] 
- R_g([X,g^{-1}\epsilon g]) \nonumber \\
&=& R_g(\delta X)\,.
\end{eqnarray}
Note that the last two terms of the first line vanish in the case of abelian $r$-matrices.
From (\ref{VOA})\,, the deformed action is invariant under the translation 
along the $x^a$ and $x^b$ directions. Then the associated Noether currents are given by 
\begin{eqnarray}
P_a^\alpha&=&-\gamma^{\alpha\beta}\Tr[\,g_X^{-1}a\,g_X\,P(J_\beta)\,]\,, \qquad 
P_b^\alpha=-\gamma^{\alpha\beta}\Tr[\,g_X^{-1}b\,g_X\,P(J_\beta)\,]\,.
\end{eqnarray}

\medskip 

One can rewrite the replacement law with these Noether currents. 
In particular, the $\sigma$-components are given by 
\begin{eqnarray}
\partial_\sigma \tilde{x}^a = \partial_\sigma x^a +2\eta\, P_b^\tau\,, \qquad 
\partial_\sigma \tilde{x}^b = \partial_\sigma x^b - 2\eta\, P_a^\tau\,. \label{421}
\end{eqnarray}
Then, by integrating the equations in (\ref{421})\,, twisted boundary conditions are obtained as  
\begin{eqnarray}
\tilde{x}^a(\tau,\,\sigma=2\pi)&=& \tilde{x}^a(\tau,\,\sigma=0)+2\eta\,P_b\,,\nonumber\\
\tilde{x}^b(\tau,\,\sigma=2\pi)&=& \tilde{x}^b(\tau,\,\sigma=0)-2\eta\,P_a \,.
\end{eqnarray}
Here $P^a$ and $P^b$ are the Noether charges defined as follows:
\begin{eqnarray}
P_a\equiv-\int^{2\pi}_0\! d \sigma \, P_a^\tau\,,\quad P_b\equiv-\int^{2\pi}_0\! d \sigma \, P_b^\tau\,.
\end{eqnarray}
These boundary conditions can be recast into a simple form, 
\begin{eqnarray}
\tilde{g}(\tau,\,\sigma=2\pi)=\exp\left[2\eta\int^{2\pi}_0\! 
d\sigma \,(P_b^\tau\, a-P_a^\tau\, b)\right]\tilde{g}(\tau,\,\sigma=0)\,,
\end{eqnarray}
with a coset representative $\tilde{g}$ for the undeformed background. 
Then the twist factor can be rewritten as follows: 
\begin{eqnarray}
2\eta(P_b^\tau a-P_a^\tau b)&=&2\eta(\Tr[g_X^{-1}bg_X\,P(J_\tau)]a
-\Tr[g_X^{-1}ag_X\,P(J_\tau)]b)\nonumber\\
&=&2\eta g R_g(J_\tau)g^{-1} = g(J_\sigma-A_\sigma)g^{-1}\nonumber\\
&\equiv&J^g_\sigma\,.
\end{eqnarray}
With this relation, the twisted boundary condition can be rewritten into a general form, 
\begin{eqnarray}
\tilde{g}(\tau,\,\sigma=2\pi)=\exp\left[\int^{2\pi}_0\! d\sigma\, 
J^g_\sigma\right]\tilde{g}(\tau,\,\sigma=0)\,, 
\end{eqnarray}
which is now independent of $x^a$ and $x^b$\,.

\subsection{The case of mCYBE}

So far, we have concentrated on the homogeneous CYBE case. 
Now it is easy to generalize the Lax pair to the mCYBE case. 
All we have to do is a slight modification to add 
an appropriate term to the previous Lax pair  (\ref{hLax}) as follows: 
\begin{eqnarray}
\label{mLax}
\mL_\pm&=&J^n_\pm+\lambda^{\pm1}(J^p_\pm+\omega \,\eta^2 J^{\tilde{k}}_\pm)\,, \qquad 
J^{\tilde{k}}_\pm \equiv J^\mu_\pm k_\mu\,.
\end{eqnarray}
Here $\omega$ is a constant that appears on the r.h.s.\ of the mCYBE in (\ref{CYBE-2})\,. 

\medskip 

Note here that the expression (\ref{mLax}) contains the case of classical $r$-matrices 
for the general $\kappa$-deformations of 4D Poincar\'e algebra \cite{kappa-Min}. 
The proof that the Lax pair (\ref{mLax}) works well is quite similar to the one in \cite{kappa-Min}. 
Hence we will not try to prove it here.

\section{Other examples}

In this section, we will present explicit forms of Lax pairs 
for twisted Minkowski spacetimes associated with classical $r$-matrices of the class (a)-1. 
We also derive twisted boundary conditions on the undeformed coordinates 
by following a simple replacement rule \cite{Frolov,AAF,KKSY}. 
It is easy to rewrite the twisted boundary conditions into simple forms 
by adopting appropriate representatives of group element. 
Hence we will work with a suitable representative for each of classical $r$-matrices,  
rather than the standard one (\ref{para})\,.

\medskip 

In the following, we will present explicit forms of Lax pairs and twisted boundary conditions 
for three cases, 1) pp-wave backgrounds, 2) the Hashimoto-Sethi background and 
3) the Spradlin-Takayanagi-Volovich background. 

\subsection{pp-wave backgrounds}

We consider an abelian classical $r$-matrix
\begin{eqnarray}
r=\frac{1}{2\sqrt{2}} \,(p_0-p_3)\wedge n_{12}\,.
\end{eqnarray}
This classical $r$-matrix is associated with pp-wave backgrounds \cite{NW}\,. 

\subsubsection*{The deformed background}

A convenient representative of group element is given by  
\begin{eqnarray}
g(\tau,\sigma)&=&\exp\left[x^-\frac{(p_0-p_3)}{\sqrt{2}}-\theta\,n_{12}\right]
\exp\left[x^+\,\frac{(p_0+p_3)}{\sqrt{2}}+r\,p_1\right]\nonumber \\
&=&\exp\left[x^+\,\frac{(p_0+p_3)}{\sqrt{2}}+x^-\frac{(p_0-p_3)}{\sqrt{2}}
+r\cos\theta\,p_1+r\sin\theta\,p_2\right]\exp[-\theta\,n_{12}]\,. \label{rep-pp}
\end{eqnarray}
Here $x^+$ and $x^-$ are the light-cone coordinates defined as 
\begin{eqnarray}
x^+ \equiv \frac{x^0+x^3}{\sqrt{2}}\,,\quad x^- \equiv \frac{x^0-x^3}{\sqrt{2}}\,.
\end{eqnarray}
The representation (\ref{rep-pp}) is equivalent to the standard one (\ref{para}) 
via a gauge transformation. 

\medskip 

Now the undeformed current $A=g^{-1}dg$ is given by 
\begin{eqnarray}
A_\pm&=&\partial_\pm x^+\,\frac{p_0+p_3}{\sqrt{2}}+\partial_\pm x^-\,
\frac{p_0-p_3}{\sqrt{2}} 
+\partial_\pm r\,p_1+\partial_\pm\theta[r\,p_2-n_{12}]\,,
\end{eqnarray}
and it leads to the standard metric of 4D Minkowski spacetime: 
\begin{eqnarray}
ds^2=-2dx^+dx^-+dr^2+r^2d\theta^2\,.
\end{eqnarray}
Then the deformed current $J_\pm$ is expanded like 
\begin{eqnarray}
J_\pm&=&\partial_\pm x^+\,\frac{p_0+p_3}{\sqrt{2}} 
+ (\partial_\pm x^-\mp\eta r^2\partial_\pm \theta+\eta^2r^2\partial_\pm x^+)\,
\frac{p_0-p_3}{\sqrt{2}} \nonumber \\
&&\qquad\quad\qquad\qquad+\partial_\pm r\,p_1
+(\partial_\pm\theta\mp\eta \partial_\pm x^+)[r\,p_2-n_{12}]\,.
\end{eqnarray}
With this deformed current, the resulting metric and NS-NS two-form are given by 
\begin{eqnarray}
ds^2 &=& -2dx^+dx^--\eta^2r^2(dx^+)^2+dr^2+r^2d\theta^2\,,\nonumber \\
B &=& \eta\, r^2 dx^+\wedge d\theta\,.
\end{eqnarray}

\subsubsection*{Lax pair}

The associated Lax pair is given by
\begin{eqnarray}
\mL_\pm&=&-(\partial_\pm\theta\mp\eta \partial_\pm x^+)\,n_{12}
+\la^{\pm1}\biggl[\partial_\pm r\,p_1
+r(\partial_\pm\theta\mp\eta \partial_\pm x^+)\,p_2 \nonumber \\
&& \quad\quad+\partial_\pm x^+\,\frac{p_0+p_3}{\sqrt{2}} 
+ (\partial_\pm x^-\mp\eta r^2\partial_\pm \theta+\eta^2r^2\partial_\pm x^+)\,
\frac{p_0-p_3}{\sqrt{2}}\biggr]\,.
\label{Null-TsT}
\end{eqnarray}
By comparing the deformed current with the undeformed one, 
one can identify the following replacement rule:
\begin{eqnarray}
\partial_\pm x^- &\to&~~ \partial_\pm x^-\mp\eta\, r^2\partial_\pm \theta
+\eta^2r^2\partial_\pm x^+\,, \nonumber \\
\partial_\pm\theta ~~&\to&~~ \partial_\pm\theta\mp\eta\, \partial_\pm x^+\,.
\end{eqnarray}
Note that the deformed Lax pair can also be reproduced 
by applying the replacement rule (\ref{Null-TsT}) 
to the undeformed Lax pair (\ref{gaugetr}) in which $U(1)$-directions are manifest.

\subsubsection*{Twisted boundary condition}

The pp-wave background with usual boundary conditions can be seen as 
as the undeformed Minkowski space with a twisted boundary conditions, as in the Melvin case. 
In terms of the undeformed group element $\tilde{g}(\tau,\sigma)$\,, 
the twisted boundary condition is given by
\begin{eqnarray}
\tilde{g}(\tau,2\pi)=\text{P}\exp\left[\int_0^{2\pi}\!\!d\sigma\,J^g_\sigma\,\right]
\exp[2\pi m(-n_{12})]\,\tilde{g}(\tau,0)\,,
\end{eqnarray}  
where the integer $m$ is a winding number for $\theta$\,.
Here the gauge-transformed current 
$J^g_\alpha=gJ_\alpha g^{-1}+g\partial_\alpha g^{-1}$ takes the following form:
\begin{eqnarray}
J_\sigma^g&=&-\eta\, r^2(\dot{\theta}-\eta x'^+)\frac{p_0-p_3}{\sqrt{2}}
+\eta\,\dot{x}^+n_{12}\nonumber \\
&=&-\eta\, \mathcal{J}^\tau_\theta\,\frac{p_0-p_3}{\sqrt{2}}-\eta\, P^\tau_-\,n_{12}\,.
\end{eqnarray}
More concretely, the twisted boundary condition can be rewritten as 
\begin{eqnarray}
\tilde{x}^-(\tau,2\pi)&=&\tilde{x}^-(\tau,0)-\eta\,\mathcal{J}_\theta\,,\nonumber \\
\tilde{\theta}^-(\tau,2\pi)&=&\tilde{\theta}^-(\tau,0)+\eta\,P_-+2\pi m\,.
\end{eqnarray}
Here $\mathcal{J}_\theta$ and $P_-$ are Noether charges for the invariance under 
the rotation and translation along the $\theta$ and $x^-$ directions, respectively.

\subsection{Hashimoto-Sethi background}

Let us next consider a classical $r$-matrix, 
\begin{eqnarray}
r=-\frac{1}{2\sqrt{2}}p_2\wedge (n_{01}+n_{13})\,.
\end{eqnarray}
This $r$-matrix leads to a time-dependent background obtained via a Melvin Null twist 
of Minkowski spacetime \cite{HS}. 

\subsubsection*{The deformed background}

A useful group parametrization is given by
\begin{eqnarray}
g(\tau,\sigma)&=&\exp\left[z\,p_2+y\,\frac{n_{01}+n_{13}}{\sqrt{2}}\right]
\exp\left[y^+\,\frac{p_0+p_3}{\sqrt{2}}+y^-\,\frac{p_0-p_3}{\sqrt{2}}\right]\nonumber \\
&=&\exp\left[y^+y\,p_1+z\,p_2+y^+\,\frac{p_0+p_3}{\sqrt{2}}+(y^-+\frac{1}{2}y^+y^2)\,
\frac{p_0-p_3}{\sqrt{2}}\right]\nonumber \\
&&\quad \times\exp\left[y\,\frac{n_{01}+n_{13}}{\sqrt{2}}\right]\,.
\end{eqnarray}
Here we have performed the following coordinate transformation: 
\begin{eqnarray}
x^1=y^+y\,,\quad x^2=z\,,\quad x^+=y^+\,,\quad x^-=y^-+\frac{1}{2}y^+y^2\,.
\end{eqnarray}
Note that this coordinate transformation is singular at $y^+=0$\,. 
Now the undeformed current $A=g^{-1}dg$ and the associated singular metric are given by, respectively,  
\begin{eqnarray}
A_\pm&=&\partial_\pm y^+\frac{p_0+p_3}{\sqrt{2}}
+\partial_\pm y^-\frac{p_0-p_3}{\sqrt{2}}\nonumber \\
&&+\partial_\pm y\left[y^+\,p_1+\frac{n_{13}+n_{01}}{\sqrt{2}}\right]+\partial_\pm z\,p_2\,, \\ 
ds^2 &=& -2dy^+dy^-+(y^+)^2dy^2+dz^2\,.
\label{HS-flat}
\end{eqnarray}
Then the deformed current $J_\pm$ is given by 
\begin{eqnarray}
J_\pm&=&\partial_\pm y^+\frac{p_0+p_3}{\sqrt{2}}+\partial_\pm y^-\frac{p_0-p_3}{\sqrt{2}}\nonumber \\
&&+G_{{\rm HS}}\,(\partial_\pm y\pm\eta \partial_\pm z)
\left[y^+\,p_1+\frac{n_{13}+n_{01}}{\sqrt{2}}\right]+G_{{\rm HS}}
\bigl[\partial_\pm z\mp\eta (y^+)^2\partial_\pm y\bigr]p_2\,,
\end{eqnarray}
with a scalar function $G_{\rm HS}$ defined as  
\begin{eqnarray}
G^{-1}_{{\rm HS}} \equiv 1+\eta^2 (y^+)^2\,.
\end{eqnarray}
As a result, the deformed metric and NS-NS two-form are obtained as 
\begin{eqnarray}
ds^2 &=& -2dy^+dy^-+G_{{\rm HS}}\left[(y^+)^2dy^2+dz^2\right]\,,\nonumber \\
B &=& \eta\, (y^+)^2 G_{{\rm HS}}\,dy\wedge dz\,.
\end{eqnarray}
This background is also realized by performing a TsT-transformation 
with the $y$ and $z$ in the undeformed background (\ref{HS-flat}) 
under the identification \cite{LMS,HS}
\begin{eqnarray}
g(\tau,\sigma)\sim g_0\,g(\tau,\sigma)\,,\qquad
g_0=\exp\left[2\pi\left(p_2+\frac{n_{01}+n_{13}}{\sqrt{2}}\right)\right]\,.
\end{eqnarray}

\subsubsection*{Lax pair}

The resulting Lax pair for the Hashimoto-Sethi background is given by
\begin{eqnarray}
\mL_\pm&=&G_{{\rm HS}}(\partial_\pm y\pm\eta \partial_\pm z)
\frac{n_{13}+n_{01}}{\sqrt{2}}\nonumber \\
&&+\la^{\pm1}\Bigl[\partial_\pm y^+\frac{p_0+p_3}{\sqrt{2}}
+\partial_\pm y^-\frac{p_0-p_3}{\sqrt{2}}\nonumber \\
&&\quad\quad+G_{{\rm HS}}\,y^+(\partial_\pm y\pm\eta \partial_\pm z)\,p_1
+G_{{\rm HS}}\left[\partial_\pm z\mp\eta (y^+)^2\partial_\pm y\right]p_2\Bigr]\,.
\end{eqnarray}
By comparing $J$ with $A$\,, one can read off the following rule: 
\begin{eqnarray}
\partial_\pm y ~~&\to&~~ G_{{\rm HS}}\,[\partial_\pm y\pm\eta \partial_\pm z]\,,\nonumber \\
\partial_\pm z ~~&\to&~~ G_{{\rm HS}}\,\left[\partial_\pm z\mp\eta (y^+)^2\partial_\pm y\right]\,.
\end{eqnarray}

\subsubsection*{Twisted boundary condition}

In the present case, the deformed background is equivalent to the undeformed background 
with the following twisted boundary condition: 
\begin{eqnarray}
\tilde{g}(\tau,2\pi)=\text{P}\exp\left[\int_0^{2\pi}\!\! d\sigma\,J^g_\sigma\,\right]
\exp\left[2\pi\left(m\,p_2+w\,\frac{n_{01}+n_{13}}{\sqrt{2}}\right)\right]\tilde{g}(\tau,0)\,,
\end{eqnarray}  
where $\tilde{g}(\tau,\sigma)$ is a group element for the undeformed background.  
The integers $m$ and $w$ are winding numbers defined as 
\begin{eqnarray}
y(\tau,2\pi)-y(\tau,0) \equiv 2\pi w\,,\qquad z(\tau,2\pi)-z(\tau,0) \equiv 2\pi m\,.
\end{eqnarray}
The gauge-transformed current $J^g_\alpha=gJ_\alpha g^{-1}+g\partial_\alpha g^{-1}$ 
is given by 
\begin{eqnarray}
J_\sigma^g 
&=& -\eta\, (y^+)^2G_{{\rm HS}}\,(\dot{y}+\eta\, z')\,p_2+\eta G_{{\rm HS}}
\left[\dot{z}-\eta\, (y^+)^2y'\right]\frac{n_{01}+n_{13}}{\sqrt{2}}\nonumber \\
&=&-\eta\,P^\tau_y\,p_2+\eta\,P^\tau_z\,\frac{n_{01}+n_{13}}{\sqrt{2}}\,.
\end{eqnarray}
More concretely, the twisted boundary condition can also be rewritten as 
\begin{eqnarray}
\tilde{y}(\tau,2\pi)&=&\tilde{y}(\tau,0)+\eta\,P_z+2\pi w\,,\nonumber \\
\tilde{z}(\tau,2\pi)&=&\tilde{z}(\tau,0)-\eta\,P_y+2\pi m\,,
\end{eqnarray}
where $P_z$ and $P_y$ are conserved charges for the translation invariance 
along the $z$ and $y$ directions, respectively.

\subsection{Spradlin-Takayanagi-Volovich background}

Finally, we shall consider an abelian classical $r$-matrix composed of two rotation generators,
\begin{eqnarray}
r=\frac{1}{2}n_{12}\wedge n_{03}\,.
\label{STV-r}
\end{eqnarray}
The corresponding background is a time-dependent background \cite{STV} as explained below. 

\subsubsection*{The deformed background}

Let us parametrize a group element like 
\begin{eqnarray}
g(\tau,\sigma)&=&\exp\left[\,\phi\,n_{03}-\theta\,n_{12}\,\right]
\exp\left[\,t\,p_0+r\,p_1\,\right]\nonumber \\
&=&\exp\left[\,t\cosh\phi\,p_0+t\sinh\phi\,p_3+r\cos\theta\,p_1
+r\sin\theta p_2\,\right] \nonumber \\
&& \times \exp[\,\phi\,n_{03}-\theta\,n_{12}\,]\,. 
\end{eqnarray}
Here we have introduced new coordinates $t$ and $\phi$ by performing a coordinate transformation  
$x^0=t\,\cosh\phi$ and $x^3=t\,\sinh\phi$\,. 
Then the undeformed current $A$ is expanded as 
\begin{eqnarray}
A_\pm &=& \partial_\pm r\,p_1+\partial_\pm t\,p_0+\partial_\pm\theta\,
[r\,p_2-n_{12}]+\partial_\pm \phi\,[t\,p_3+n_{03}]\,.
\end{eqnarray}
As a result, the associated undeformed metric takes the form
\begin{eqnarray}
ds^2=-dt^2+t^2d\phi^2+dr^2+r^2d\theta^2\,.
\end{eqnarray}
Then the deformed current $J_\pm$ with the classical $r$-matrix (\ref{STV-r}) is evaluated as 
\begin{eqnarray}
J_\pm&=&\partial_\pm r\,p_1+\partial_\pm t\,p_0\nonumber \\
&&+G_{{\rm STV}}\,(\partial_\pm\theta\mp\eta t^2\partial_\pm\phi)\,\bigl(r\,p_2-n_{12} \bigr) 
\nonumber \\
&&+G_{{\rm STV}}\,(\partial_\pm \phi\pm\eta r^2\partial_\pm\theta)\, \bigl(t\,p_3+n_{03}\bigr)\,,
\end{eqnarray}
where we have introduced a scalar function $G_{\rm STV}$ defined as 
\begin{eqnarray}
G^{-1}_{{\rm STV}} \equiv 1+\eta^2 r^2 t^2\,.
\end{eqnarray}
The resulting deformed background is therefore given by 
\begin{eqnarray}
ds^2 &=& -dt^2+dr^2+G_{{\rm STV}}\,(t^2d\phi^2+r^2d\theta^2)\,,\nonumber \\
B &=& \eta\, t^2r^2\, G_{{\rm STV}}\,d\phi\wedge d\theta\,. 
\end{eqnarray}

\subsubsection*{Lax pair}

The associated Lax pair with a spectral parameter $\lambda$ is expressed as  
\begin{eqnarray}
\mL_\pm&=&-G_{{\rm STV}}\,(\partial_\pm\theta\mp\eta\, t^2\partial_\pm\phi)\,n_{12}+G_{{\rm STV}}
(\partial_\pm \phi\pm\eta\, r^2\partial_\pm\theta)\,n_{03}\nonumber \\
&&+\la^{\pm1}\Bigl[\partial_\pm r\,p_1+\partial_\pm t\,p_0\nonumber \\
&&\quad\quad+r\,G_{{\rm STV}}\,(\partial_\pm\theta\mp\eta\, t^2\partial_\pm\phi)\,p_2+t\,G_{{\rm STV}}\,
(\partial_\pm \phi\pm\eta\, r^2\partial_\pm\theta)\,p_3\Bigr]\,.
\end{eqnarray}

\subsubsection*{Twisted boundary condition}

Similarly, by comparing the deformed current with the undeformed one, 
one can read off the following replacement rule:  
\begin{eqnarray}
\partial_\pm\theta ~~&\rightarrow&~~ 
G_{{\rm STV}}\,(\partial_\pm\theta\mp\eta\, t^2\partial_\pm\phi)\,,\nonumber \\
\partial_\pm \phi ~~&\rightarrow&~~ 
G_{{\rm STV}}\,(\partial_\pm \phi\pm\eta\, r^2\partial_\pm\theta)\,.
\end{eqnarray}
This rule implies a twisted boundary condition for the undeformed coordinates, 
\begin{eqnarray}
\tilde{g}(\tau,2\pi)=\text{P}\exp\left[\int_0^{2\pi}\!\!d\sigma\,
J^g_\sigma\,\right]
\exp\left[2\pi\left(m\,n_{03}-w\,n_{12}\right)\right]\tilde{g}(\tau,0)\,. 
\end{eqnarray}
Here the integer numbers $w$ and $m$ are winding numbers for the angles $\theta$ and $\phi$\,.
The gauge-transformed current $J^g_\alpha=gJ_\alpha g^{-1}+g\partial_\alpha g^{-1}$ is given by 
\begin{eqnarray}
J_\sigma^g &=& \eta\, r^2\,G_{{\rm STV}}(\dot{\theta}-\eta\, t^2\phi')n_{03}
+\eta\, t^2\,G_{{\rm STV}}(\dot{\phi}+\eta\, r^2\theta')n_{12}\nonumber \\
&=& \eta\, \mathcal{J}^\tau_\theta\,n_{03} + \eta\, \mathcal{J}^\tau_\phi\,n_{12}
\,.
\end{eqnarray}
This expression can also be recast in 
\begin{eqnarray}
\tilde{\theta}(\tau,2\pi)&=&\tilde{\theta}(\tau,0)-\eta\, \mathcal{J}_\phi+2\pi\,w\,,\nonumber \\
\tilde{\phi}(\tau,2\pi)&=&\tilde{\phi}(\tau,0)+\eta\, \mathcal{J}_\theta+2\pi\,m\,,
\end{eqnarray}
where $\mathcal{J}_\phi$ and $\mathcal{J}_\theta$ are Noether charges 
for the rotational invariance for the $\phi$ and $\theta$ directions.

\section{Conclusion and discussion}

In this paper, we have proceeded to study Yang-Baxter deformations of 4D Minkowski spacetime 
based on a conformal embedding. In the first place, we have revisited a Melvin background 
and argued a Lax pair by adopting a simple replacement law invented 
in \cite{KKSY}. Due to this argument, one can readily deduce a general expression of Lax pair. 
Then the anticipated Lax pair has been shown to work for arbitrary classical $r$-matrices 
with Poincar\'e generators. As other examples, we have presented Lax pairs 
for pp-wave backgrounds, the Hashimoto-Sethi background, 
the Spradlin-Takayanagi-Volovich background. 

\medskip 

There are a lot of open problems. We have considered only the bosonic sector so far. 
It would be very interesting to investigate the RR-sector and dilaton by including spacetime fermions 
(c.f., see \cite{super}). In fact, it is much easier to perform a supercoset construction 
for Yang-Baxter deformed Minkowski spacetimes, 
in comparison to the deformed AdS$_5\times$S$^5$\,.  
It is also nice to study non-local charges obtained from the monodromy matrices  
constructed from the Lax pair derived here. 
The resulting algebras may be related to the preceding results obtained in \cite{HM}. 

\medskip 

As another direction, it would be of interest to consider 
Yang-Baxter deformations of other backgrounds such as pp-wave backgrounds 
and Lifshitz spacetimes \cite{KLM}. Along this direction, it has been shown that 
the Nappi-Witten background \cite{NW} is Yang-Baxter invariant 
at the quantum level \cite{KY-NW}. Due to the argument in \cite{SYY}, 
Lifshitz spacetimes would be invariant as well. It is very nice to 
check whether this anticipation is valid or not.

\subsection*{Acknowledgments}

We are very grateful to Andrzej Borowiec, Jerzy Lukierski,  Takuya Matsumoto, 
Domenico Orlando, Susanne Reffert and Stijn van Tongeren 
for useful discussions and collaborations on the previous works. 
The work of K.Y. is supported by Supporting Program for Interaction-based Initiative Team Studies 
(SPIRITS) from Kyoto University and by the JSPS Grant-in-Aid for Scientific Research (C) No.15K05051.
This work is also supported in part by the JSPS Japan-Russia Research Cooperative Program 
and the JSPS Japan-Hungary Research Cooperative Program.

\appendix 

\section*{Appendix}

\section{Notation and convention}

We will summarize here our notation and convention of the $\alg{so}(2,4)$ generators. 

\medskip

In this paper, we use the following basis of $\alg{su}(2,2)\simeq\alg{so}(2,4)$~: 
\begin{eqnarray}
\alg{su}(2,2)=\text{span}_{\mathbb{R}}
\left\{~\ga_\mu\,,\ga_5\,,n_{\mu\nu}=\frac{1}{4}[\ga_\mu\,,\ga_\nu]\,, 
n_{\mu5}=\frac{1}{4}[\ga_\mu\,,\ga_5]~|
~~\mu,\nu=0,1,2,3~\right\}\,.
\end{eqnarray}
Here $\gamma_{\mu}$'s are gamma matrices satisfying the Dirac algebra:
\begin{eqnarray}
\{\gamma_{\mu}\,, \gamma_{\nu}\} = 2\eta_{\mu\nu}\,. 
\end{eqnarray}
Here $\eta_{\mu\nu}$ is the standard Minkowski metric with mostly plus. 
It is convenient to adopt the following matrix realization of $\gamma_{\mu}$'s, 
\begin{eqnarray}
&& \gamma_1=
\begin{pmatrix}
\;0~&~0~&~0~&-1\\
0&0&1&~0\\
0&1&0&~0\\
-1&0&0&~0\\
\end{pmatrix}\,, 
\quad 
\gamma_2=
\begin{pmatrix}
\;0~&~0~&~0~&~i\\
0&0&i&~0\\
0&-i&0&~0\\
-i&0&0&~0\\
\end{pmatrix}\,, \quad 
\gamma_3=
\begin{pmatrix}
\;0~&~0~&~1~&~0\\
0&0&0&~1\\
1&0&0&~0\\
0&1&0&~0\\
\end{pmatrix}\,,\nonumber \\ 
&& \gamma_0=i\gamma_4=
\begin{pmatrix}
\;0~&~0~&1~&0\\
0&0&0&-1\\
-1&0&0&~0\\
0&1&0&~0\\
\end{pmatrix}\,, \quad 
\gamma_5=i\gamma_1\gamma_2\gamma_3\gamma_0=
\begin{pmatrix}
\;1~&~0~&~0~&0\\
0&1&0&~0\\
0&0&-1&~0\\
0&0&0&-1\\
\end{pmatrix}\,.
\end{eqnarray}

\medskip

It is also helpful to use the conformal basis for $\mathfrak{so}(2,4)$~: 
\begin{eqnarray}
\mathfrak{so}(2,4) = {\rm span}_{\mathbb{R}}\{~p_{\mu}\,, n_{\mu\nu}\,, \hat{d}\,, k_{\mu}~|
~~\mu,\nu=0,1,2,3~\}\,.
\label{cb}
\end{eqnarray}
Here the translation generators $p_{\mu}$\,, 
the Lorentz rotation generators $n_{\mu\nu}$\,, the dilatation $\hat{d}$  
and the special conformal generators $k_{\mu}$ 
are represented by, respectively,  
\begin{align}
p_{\mu} \equiv \frac{1}{2}(\ga_{\mu}-2n_{\mu5})\,,  \quad 
\hat{d} \equiv \frac{1}{2}\gamma_5\,,  
\quad 
k_{\mu} \equiv \frac{1}{2}(\ga_{\mu}+2n_{\mu5})\,. 
\end{align}
The non-vanishing commutation relations are given by 
\begin{eqnarray}
[p_\mu ,k_\nu]&=& 2 (n_{\mu\nu}+\eta_{\mu\nu}\, \hat{d}\,)\,,\quad 
[\hat{d},p_{\mu}]=p_\mu\,,\quad [\hat{d},k_\mu]=-k_\mu\,,\nonumber\\
\left[p_\mu,n_{\nu\rho}\right] &=&\eta_{\mu\nu}\, p_\rho-\eta_{\mu\rho}\, p_\nu \,,\quad 
[k_\mu,n_{\nu\rho}]=\eta_{\mu\nu}\,k_\rho-\eta_{\mu\rho}\,k_\nu\,, \nonumber\\
\left[n_{\mu\nu} ,n_{\rho\sigma}\right]&=&\eta_{\mu\sigma}\, n_{\nu\rho}
+\eta_{\nu\rho}\,n_{\mu\sigma}-\eta_{\mu\rho}\,n_{\nu\sigma}-\eta_{\nu\sigma}\,n_{\mu\rho}\,.
\end{eqnarray}

\section{A conformal embedding of 4D Minkowski spacetime}

To introduce Yang-Baxter deformations of 4D Minkowski spacetime, 
it is necessary to perform a coset construction of flat space. 
However there is an obstacle that the Killing form on the standard coset 
$ISO(1,3)/SO(1,3)$ is degenerate. Hence, in order to avoid this issue, 
we will embed this coset into a conformal group $SO(2,4)$\,, as explained below.

\medskip 

A representative of group element $g$ is given by 
\begin{equation}
g=\exp\Bigl[ \,p_0\, x^0 + p_1\, x^1 + p_2\, x^2 + p_3\, x^3 \,\Bigr] \,.
\label{para}
\end{equation}
Note that this parametrization may be interpreted as a slice of Poincar\'e AdS$_5$\,, 
in which the representative is given by $g=\exp[x^\mu p_\mu]\exp[\hat{d}\log z]$ 
with the AdS radial direcrion $z$\,. 

\medskip 

The left-invariant one-form $A = g^{-1}dg$ can be expanded as 
\begin{eqnarray}
A \equiv e^{\mu} p_{\mu}\,, \qquad
e^{\mu} = dx^{\mu}\,. 
\end{eqnarray}
Then, in terms of the vierbeins $e^{\mu}$\,, the metric of Minkowski spacetime is given by
\begin{eqnarray}
ds^2 = \eta_{\mu\nu}e^{\mu}e^{\nu} = \eta_{\mu\nu}dx^{\mu}dx^{\nu}\,. 
\label{metric-P}
\end{eqnarray}
By using the relations
\begin{eqnarray}
e^\mu=\frac{1}{2}\Tr(\ga^\mu A)\,,
\end{eqnarray}
the metric (\ref{metric-P}) can be recast into the following form: 
\begin{eqnarray}
ds^2 = \eta_{\mu\nu}e^{\mu}e^{\nu} = \Tr(AP(A))\,.  
\label{metric-new}
\end{eqnarray}
Here we have introduced a projection operator defined as 
\begin{align}
P(x) \equiv \frac{1}{4} \Bigl[-\ga_{0}\,\Tr(\ga_{0}\,x) + \sum_{i=1}^3\ga_{i}\, \Tr(\ga_{i}\,x) \Bigr]\,.
\label{Proj}
\end{align}
Note that $\ga^\mu$'s contained in (\ref{Proj}) are elements of 4D conformal algebra $\alg{so}(2,4)$ 
rather than 4D Poincar\'e algebra $\alg{iso}(1,3)$\,.
Thus the projection (\ref{Proj}) implies that our coset construction of 4D Minkowski spacetime 
is realized as a slice of Poincar\'e AdS$_5$\,. 

\medskip

As a side note, it is straightforward to generalize the projection operator 
to the general symmetric two-form \cite{NW,SYY}. Due to this observation, 
Yang-Baxter deformations can be applied to the Schr\"odinger and Lifshitz cosets argued in \cite{SYY}. 
For a recent progress along this line, see the work \cite{KY-NW} in which 
Yang-Baxter deformations of the Nappi-Witten model have been studied.

\end{document}